\documentclass[letterpaper,12pt]{article}

\usepackage{osajnl2}
\usepackage{amsfonts, bbm, amsfonts}
\usepackage[numbers,sort&compress]{natbib}
\usepackage{graphicx}

\usepackage{epstopdf}

\begin{document}

\title{A posterior quantum dynamics for a continuous diffusion
observation of a coherent channel}

\author{Anita D\c{a}browska$^{*}$ and Przemys{\l}aw Staszewski}
\address{Department of Theoretical Foundations of Biomedical Sciences
and Medical Informatics\\ Collegium Medicum, Nicolaus Copernicus
University\\  ul.~Jagiello\'nska 15, 85-067 Bydgoszcz, Poland }
\address{$^*$Corresponding author: adabro@cm.umk.pl}

\begin{abstract}
We present the Belavkin filtering equation for the
intense balanced heterodyne detection in a unitary model of an
indirect observation. The measuring apparatus modelled by a Bose
field is initially prepared in a coherent state and the observed
process is a diffusion one. We prove that this filtering equation
is relaxing: any initial square-integrable function tends
asymptotically to a coherent state with an amplitude depending on
the coupling constant and the initial state of the apparatus. The
time-development of a squeezed coherent state is studied and
compared with the previous results obtained for the measuring
apparatus prepared initially in the vacuum state.
\end{abstract}


\maketitle
\section{Introduction}
The theory of quantum measurements continuous in time is firmly
based on quantum stochastic calculus (QSC) developed by Hudson and
Parthasarathy \cite{HudPar84, Par92}. The time development of the
posterior state conditioned by a trajectory of the results of a
continuous measurement is given by the Belavkin filtering equation
\cite{Bel99, Bel90, BarBel91, Bel2002}. The measurement is taken on
a Bose field interacting with the quantum system in question and
enables one to perform its indirect observation. The Bose field can
be treated as an approximation to the electromagnetic field. The
filtering equation has the form of the Ito quantum stochastic
differential equation and plays the r\^{o}le analogous to that of
the Schr\"odinger equation for an unobserved quantum systems. The
observed process has the properties of a diffusion or/and a
counting one. The Belavkin filtering equation was obtained under
the assumption that the Bose field modelling the apparatus was
initially prepared in the vacuum state. In \cite{BarPag96,
GoughK, DSOSAB2011} this assumption has been relaxed and the
filtering equation has been derived for the Bose field prepared in
a coherent state for the counting and diffusion observations.

The aim of this paper is twofold --- we present the Belavkin
equation for the diffusion observation with the apparatus prepared
initially in a coherent state and we discuss the time development
of a squeezed coherent state undergoing a balance heterodyne measurement
\cite{GK05}.
In contrast to the result of \cite{DSPLA2011}, showing that for the apparatus
initially prepared in the vacuum state a single and double
heterodyne detection does not destroy the squeezed coherent state
and drives the system asymptotically to the vacuum state, the
state asymptotically relaxes to the coherent
 one with the amplitude independent of the initial state of the system.
Next, we generalize this observation and prove that any
square-integrable initial wave function relaxes to the coherent
state with the given amplitude depending on the strength of the
interaction between the system (single-mode field) and the
apparatus and the initial state of the apparatus. Consequently,
the considered filtering equation describes in fact a control of
the quantum system by driving its state to the coherent one with
the given amplitude.

Though the methods of QSC are neither widely used nor well-known by
physicists, they deserve in our opinion more interest as the
effective tools in modelling physical systems interacting with
measuring devices. We refer the readers who not familiar with the
theory of quantum measurements continuous in time to the books
\cite{Car93, DM04, WM10} or the recent papers \cite{Bru02,
JacSte06, HSM05, BHJ07}. More rigorous approach to the subject can
be found in \cite{Bar06, BG06, BarGre09, GarZol10}. The
experimental achievements in the area of continuous in time
observations of quantum systems were reported, for instance, in
\cite{MabKim99, Arm02,GSDM03, GSM06}.

The paper is organized as follows. In Section 2 we present the
basic rules of quantum stochastic calculus. Section 3 is devoted to
the presentation of a linear version of the Belavkin filtering
equation for a balanced heterodyne detection of the diffusion type
for the apparatus prepared initially in the coherent state. We
choose the linear form of the filtering equation (for an
unnormalized posterior wave function) instead of the nonlinear
version derived in \cite{DSPLA2011}, because it is more convenient
to deal with. The physical interpretation of the linear filtering
equation one can find, for instance, in \cite{GoeGra94, BarGre09}.
We put forward here the approach of the generating map for the
underlined continuous observation. In Section 4 we discuss the time
development of a coherent and a squeezed coherent state evolving
under a continuous diffusion observation of a coherent channel.

\section{Quantum stochastic calculus}

In this section we recall some basic rules of quantum stochastic
calculus (QSC) in the boson Fock space \cite{HudPar84, Par92}.
Denote by $\mathcal{F}$ the (symmetrical) Fock space over the
Hilbert space $\mathcal{K}=\mathbb{C}^{n}\otimes
L^{2}(\mathbb{R}_{+})$ of all square integrable functions from
$\mathbb{R}_{+}$ into $\mathbb{C}^{n}$. For any $f\in \mathcal{K}$
one can define a coherent vector, $e(f)$, by the formula
%------------- --------------------------------------------------------
\begin{equation}
e(f) \;=\; \exp\left( -\frac{1}{2}
||f||_{\mathcal{K}}^{2}\right)\left(1,f,(2!)^{-1/2}f\otimes
f,(3!)^{-1/2}f\otimes f\otimes f,\ldots\right)\,.
\end{equation}
%----------------------------------------------------------------------
In particular, $e(0) =(1,0,0,\ldots) \in \mathcal{F}$ is the Fock
vacuum. The annihilation, creation and number processes:
$B_{j}(t)$, $B_{j}^{\dagger}(t)$ and $\mathit{\Lambda}_{ij}(t)$
are defined on the dense in $\mathcal{F}$ linear span of all
coherent vectors \cite{HudPar84, Par92} as follows:
%----------------------------------------------------------------------
\begin{equation}\label{basic1}
B_{j}(t)e(f) \;=\; \int_{0}^{t}f_{j}(s){\mathrm d}s\ e(f)\,,
\end{equation}
\begin{equation}\label{basic2}
\langle e(g)|B_{j}^{\dagger}(t)e(f)\rangle \;=\; \int_{0}^{t}
\overline{g_{j}(s)}{\mathrm d}s\,\langle e(g)|e(f)\rangle\,,
\end{equation}
\begin{equation}\label{basic3}
\langle e(g)|\mathit{\Lambda}_{ij}(t)e(f)\rangle\;=\;
\int_{0}^{t}\overline{g_{i}(s)}f_{j}(s)\mathrm{d}s\,\langle e(g)|
e(f)\rangle\,.
\end{equation}
%----------------------------------------------------------------------
These are the underlying processes for stochastic differential
equations (QSDEs) of the Ito type:
%----------------------------------------------------------------------
\begin{equation}\label{QSC}
\mathrm{d}M(t)\;=\;\sum_{j=1}^{n}\left(\sum_{i=1}^{n} F_{ji}(t)
\,\mathrm{d} \mathit{\Lambda}_{ji}(t)+E_{j}(t)\,
\mathrm{d}B_{j}(t)+D_{j}(t)\,
\mathrm{d}B_{j}^{\dagger}(t)\right)+C(t)\,\mathrm{d}t\,.
\end{equation}
%----------------------------------------------------------------------
In (\ref{QSC}) all the processes appearing at the Ito differentials
are adapted processes on $\mathcal{H}\otimes \mathcal{F}$, i.e.
they depend on the processes (\ref{basic1}-\ref{basic3}) up to $t$
(present instant) and commute with the Ito differentials that
``point to the future''. If $M^{\prime}(t)$ is the process which
satisfies an equation of the type (\ref{QSC}), then the
differential of the product $M(t)M^{\prime}(t)$ is given by the
formula \cite{HudPar84, Par92}
%-----------------------------------------------------------------------
\begin{equation}\label{Itofor}
{\mathrm d} \big(M(t)M^{\prime}(t)\big)= {\mathrm d}M(t)M^{\prime}(t)
+M(t) {\mathrm d}M^{\prime}(t)+{\mathrm d}M(t){\mathrm d}
M^{\prime}(t)\,.
\end{equation}
%-----------------------------------------------------------------------
The term ${\mathrm d}M(t)\,{\mathrm d}M^{\prime}(t)$ can be
computed with the help of the multiplication table:
%-------------------------------------------------------------------------
$${\mathrm d}B_{i}(t)\,{\mathrm d}B_{j}^{\dagger}(t)\;=\;\delta_{ij}\,
{\mathrm d}t\,,\;\;\; {\mathrm d}B_{i}(t)\,{\mathrm d}\mathit{\Lambda}_{kj}(t)
\;=\;\delta_{ik}\,{\mathrm d}B_{j}(t)\,,$$
\begin{equation}\label{Itotable}
{\mathrm d}\mathit{\Lambda}_{kj}(t)\,{\mathrm d}B_{i}^{\dagger}(t)
\;=\;\delta_{ji}\,
{\mathrm d}B_{k}^{\dagger}(t)\,,\;\;\; {\mathrm d}\mathit{\Lambda}_{ij}(t)\,
{\mathrm d}\mathit{\Lambda}_{kl}(t)\;=\;\delta_{jk}\,{\mathrm d}
\mathit{\Lambda}_{il}(t)\,,
\end{equation}
%-------------------------------------------------------------------------
and all other products vanish.

\section{Linear filtering equation for a balanced heterodyne scheme}

Let us consider a harmonic oscillator (system $\mathcal{S}$)
interacting with an environment modelled by the Bose field in a
coherent state $e(f)$. We assume that the unitary evolution
operator $U(t)$ of the compound system (system $\mathcal{S}$ plus
one-dimensional Bose field) satisfies the QSDE \cite{Bar06}:
%----------------------------------------------------------------------
\begin{equation}\label{unitary}
\mathrm{d}U(t)\;=\;\left[\sqrt{\mu}\ a\ \mathrm{d}B^{\dagger}(t)
-\sqrt{\mu}\   a^{\dagger}\ \mathrm{d}B(t)-
{\mu \over 2}\
a^{\dagger}a\ \mathrm{d}t-{\mathrm{i}\over \hbar }
H\mathrm{d}t\right]U(t)\,, \;\;\; U(0)\;=\;I\,,
\end{equation}
%----------------------------------------------------------------------
where $H=\hbar \omega\left(a^{\dagger}a+{1\over 2}\right)$ is the
Hamiltonian of $\mathcal{S}$, $a$ is an annihilation operator, and
$\mu \in \mathbb{R}_{+}$ stands for a real coupling constant.
Eq.~(\ref{unitary}) is written in the interaction picture with
respect to the free dynamics of the Bose field. The description of
physical assumptions leading to this evolution can be found, for
instance, in \cite{GarCol85, Bar06}. In short, the coupling is
linear in the field operators, the rotating-wave approximation
(RWA) is made, the coupling constant are independent of frequency,
and the spectrum of the reservoir is flat and broad. These
assumptions are often made in quantum optics.

\noindent Though the Bose field disturbs the free evolution of
$\mathcal{S}$, it also enables an indirect observation of
$\mathcal{S}$ continuous in time. The input processes $B(t)$,
$B^{\dagger}(t)$ refer to the field before its interaction with
$\mathcal{S}$, whereas the output processes
$B^{\mathrm{out}}(t)=U^{\dagger}(t)B(t)U(t)$,
$B^{\mathrm{out}\dagger}(t)=U^{\dagger}(t)B^{\dagger}(t)U(t)$
describe the field after the interaction with $\mathcal{S}$.

In a balance heterodyne measurement depicted in Fig.~1 the output
field, escaping from the cavity, is mixed with a strong laser
field $B_{\mathrm{lo}}(t)$ (local oscillator).
We assume that this auxiliary field, which does not interact with $\mathcal{S}$,
is initially in a coherent state $e(f_{\mathrm{lo}})$ \cite{Bar02, Bar06}.
In the paper we consider the filtering equation corresponding to the
observation of the difference of photocurrents generated by the detectors
monitoring the fields:
%----------------------------------------------------------------------
\begin{equation}
B_{1}(t)\;=\;\frac{1}{\sqrt{2}}\left(B^{\mathrm{out}}(t)+
B_{\mathrm{lo}}(t)\right)\,,\;\;\;\;\;\;\; B_{2}(t)\;=\;
\frac{1}{\sqrt{2}}\left(B^{\mathrm{out}}(t)-B_{\mathrm{lo}}(t)\right)\,.
\end{equation}
%----------------------------------------------------------------------
To derive the linear stochastic equation one can use, for example,
the Belavkin method of generating functional \cite{Bel90}. The
generating map, $\mathrm{g}(k,t)$, defined by \cite{Bel90, Bar06}
%----------------------------------------------------------------------
$$\mathrm{g}(k,t): Z\rightarrow \mathrm{g}(k,t)[Z]\,,$$
\begin{equation}\label{genmap}
\langle\psi |g(k,t)[Z]\psi\rangle\;=\;\langle\psi\otimes e(f)|
\mathrm{G}^{\mathrm{out}}(k,t)Z_{t}\psi\otimes e(f)\rangle\,,
\end{equation}
%----------------------------------------------------------------------
where
%----------------------------------------------------------------------
\begin{equation}\label{opergen}
\mathrm{G}^{\mathrm{out}}(k,t)\;=\; \langle
e(f_{\mathrm{lo}})|\exp \bigg\{ \int\limits_{0}^{t}\varepsilon\,
k(t^{\prime})\bigg( \mathrm{d}\mathit{\Lambda}_{11}(t^{\prime})
-\mathrm{d}\mathit{\Lambda}_{22}(t^{\prime}) \bigg) \bigg\}
|e(f_{\mathrm{lo}})\rangle\,,
\end{equation}
%---------------------------------------------------------------------
completely determines the observed process up to time $t$.
Here  $Z_{t}=U^{\dagger}(t)ZU(t)$ is the Heisenberg operator of
$\mathcal{S}$, $\psi$ stands for the initial state of
$\mathcal{S}$, $k$ is any integrable $c$-valued function, and
$\varepsilon^{-1}=|\,f_{\mathrm{lo}}\,|$. In the limit
$\varepsilon\rightarrow 0$ of a very intense local oscillator
field, the formula (\ref{opergen}) takes the form
%---------------------------------------------------------------------
\begin{equation}\label{output}
\mathrm{G}^{\mathrm{out}}(k,t)\;=\; \exp\bigg\{\int\limits_{0}^{t}
k(t^{\prime})\, \mathrm{d}\mathcal{Q}^{\mathrm{out}}(t^{\prime})
\bigg\}\,,
\end{equation}
%---------------------------------------------------------------------
where the output process $\mathcal{Q}^{\mathrm{out}}$ reads
%---------------------------------------------------------------------
\begin{equation}\label{mea}
\mathcal{Q}^{\mathrm{out}}(t)\;=\;\int\limits_{0}^{t}
\bigg(\mathrm{e}^{\mathrm{i}\phi(t^{\prime})}\mathrm{d}
B^{\mathrm{out}\dagger}(t^{\prime}) +\mathrm{e}^{-\mathrm{i}
\phi(t^{\prime})} \mathrm{d}B^{\mathrm{out}}(t^{\prime}) \bigg)\,,
\end{equation}
%---------------------------------------------------------------------
$\phi(t)=\arg f_{\mathrm{lo}}(t)$. By using (\ref{unitary}) and
the Ito formula (\ref{Itofor}) one obtains
%----------------------------------------------------------------------
\begin{equation}
\mathrm{d}\mathcal{Q}^{\mathrm{out}}(t)\;=\;\mathrm{e}^{\mathrm{i}
\phi(t)}
\mathrm{d} B^{\dagger}(t) +\mathrm{e}^{-\mathrm{i}\phi(t)}
\mathrm{d}B(t)+ \sqrt{\mu}\,\big(\mathrm{e}^{\mathrm{i}
\phi(t)}a_{t}^{\dagger} +\mathrm{e}^{-\mathrm{i}\phi(t)}
a_{t}\big)\mathrm{d}t\,
\end{equation}
%----------------------------------------------------------------------
and $(\mathrm{d}\mathcal{Q}^{\mathrm{out}}(t))^{2} =\mathrm{d}t$.
An explicit expression for the generating map (\ref{genmap}) one
can find by solving the differential equation for
$\mathrm{g}(k,t)$. Using the method described in \cite{Bel90}, one
can check that the generating map $\mathrm{g}(k,t)$ satisfies the
equation
%----------------------------------------------------------------------
\begin{eqnarray}\label{diff3}
\frac{\mathrm{d}}{\mathrm{d}t}\mathrm{g}(k,t)[Z] &=& \mathrm{g}(k,t)
\left[ -(K+\sqrt{\mu}\,a^{\dagger}f(t)-\sqrt{\mu}\,
a\overline{f(t)})^{\dagger}Z+\mu a^{\dagger}Za\right.\nonumber\\
&&-Z(K+\sqrt{\mu}\,a^{\dagger}f(t)-\sqrt{\mu}\,a\overline{f(t)})
+\frac{1}{2}k^{2}(t)Z\nonumber\\
&&\left.+k(t)\big(\sqrt{\mu}a^{\dagger}+\overline{f(t)}\big)
\mathrm{e}^{\mathrm{i}\phi(t)}Z+ k(t)Z\big(\sqrt{\mu}a+f(t)\big)
\mathrm{e}^{-\mathrm{i}\phi(t)}\right]\,
\end{eqnarray}
%----------------------------------------------------------------------
with $\mathrm{g}(k,0)[Z]=Z$ and $K=\frac{\mathrm{i}}{\hbar}H+
\frac{\mu}{2}a^{\dagger}a$. The solution to the Eq. (\ref{diff3})
can be written in the form \cite{Bel90}
%-------------------------------------------------------------------------
\begin{equation}\label{sol1}
\mathrm{g}(k,t)[Z] \;=\;
\int\limits_{\Omega^{t}}\mathrm{G}(k,q^{t})
V^{\dagger}(q^{t})ZV(q^{t}) \, \mathrm{d}\nu (q^{t})\,,
\end{equation}
%-------------------------------------------------------------------------
where $\nu$ is the probabilistic measure on the set $\Omega$
consisting of the continuous trajectories\break $q = [ q(t)| t>0
]$ of the observed process $\mathcal{Q}^{\mathrm{out}}(t)$,
restricted to the set $\Omega^{t}=\{q^{t}|q\in \Omega\}$ of the
trajectories $q^{t}=[q(r)|r\leq t]$ up to $t$. The stochastic
propagator $\widehat{V}(t)(q^{t})=V(q^{t})$ satisfies the
stochastic differential equation
%-------------------------------------------------------------------------
\begin{equation}\label{diff4}
\mathrm{d}\widehat{V}(t)\;=\;
-\bigg(K+\sqrt{\mu}\,a^{\dagger}f(t)+ \sqrt{\mu}\,a\,
\mathrm{e}^{-2\mathrm{i}\phi(t)}f(t)\bigg)
\widehat{V}(t)\,\mathrm{d}t+\sqrt{\mu}\,a\,
\mathrm{e}^{-\mathrm{i}\phi(t)}\widehat{V}(t)\,\mathrm{d}
\mathcal{Q}(t)\,,\;\;\;\; \widehat{V}(0)\;=\;I\,
\end{equation}
%-------------------------------------------------------------------------
and
%-------------------------------------------------------------------------
\begin{equation}
\mathrm{G}(k,q^{t})\;=\;\mathrm{G}^{\mathrm{out}}(k,t)(q)\;=\;
\exp\bigg\{\int_{0}^{t}k(t^{\prime})\,
\mathrm{d}q(t^{\prime})\bigg\}\,.
\end{equation}
%-------------------------------------------------------------------------
Hence, the posterior unnormalized wave function
$\widehat{\psi}(t)= \widehat{V}(t)\psi$ of the system
$\mathcal{S}$ satisfies the stochastic dissipative differential
equation of the form
%-------------------------------------------------------------------------
\begin{equation}\label{diff7}
\mathrm{d}\widehat{\psi}(t)\;=\;
-\bigg(K+\sqrt{\mu}\,\left(a^{\dagger}f(t)-a\,
\overline{f(t)}\right)\bigg)\widehat{\psi}(t)
\mathrm{d}t+\sqrt{\mu}\,a\, \mathrm{e}^{-\mathrm{i}\phi(t)}
\widehat{\psi}(t) \, \mathrm{d}W(t)\,,\;\;\; \widehat{\psi}(0)
\;=\;\psi\,,
\end{equation}
%-------------------------------------------------------------------------
where
%-------------------------------------------------------------------------
\begin{equation}
\mathrm{d}W(t)\;=\;\mathrm{d}\mathcal{Q}(t)-
2\,\mathrm{Re}\left(\mathrm{e}^{-\mathrm{i}\phi(t)}f(t)\right)\,
\mathrm{d}t
\end{equation}
%-------------------------------------------------------------------------
and $W(t)$ is isomorphic to the standard Wiener process.
Let us notice that the integral representation  (\ref{sol1}) of
the generating map,
%-------------------------------------------------------------------------
\begin{equation}\label{mean}
\langle \mathrm{G}^{\mathrm{out}}(k,t)Z_{t} \rangle \;=\;
\int\limits_{\Omega^{t}}\mathrm{G}(k,q^{t})\langle V(q^{t})
\psi|ZV(q^{t})\psi\rangle \,\mathrm{d}\nu (q^{t})\,,
\end{equation}
%-------------------------------------------------------------------------
gives for $Z=I$ the mean value of the output process
(\ref{output}) as the generating function of the output
probability measure
%-------------------------------------------------------------------------
\begin{equation}\label{measure}
\mathrm{d}\zeta(q^{t})\;=\;\langle V(q^{t}) \psi|V(q^{t})\psi
\rangle \, \mathrm{d}\nu(q^{t})\,.
\end{equation}
%-------------------------------------------------------------------------
Therefore, the posterior wave function $\psi(t)$ satisfying the
filtering equation (\ref{diff7}) is normalized to the probability
density $\langle V(q^{t}) \psi|V(q^{t})\psi \rangle$ of the
observed process with respect to the probability measure of the
input process. From (\ref{mean}) we obtain the posterior mean value $\langle
Z\rangle(q^{t})$ as
\begin{equation}
\langle Z\rangle(q^{t})\;=\;\langle \varphi(q^{t})|Z\varphi(q^{t})
\rangle\,,
\end{equation}
where $\widehat{\varphi}(t)(q)=\varphi(q^{t})$ and
$\widehat{\varphi}(t)= \langle \widehat{\psi}(t)|\widehat{\psi}(t)
\rangle^{-1/2}\widehat{\psi}(t)$. For the normalized posterior
wave function $\widehat{\varphi}(t)$ we get the nonlinear filtering
equation
%-------------------------------------------------------------------------
\begin{eqnarray}\label{diff8}
\lefteqn{\mathrm{d}\widehat{\varphi}(t)\;=\;-\left(K+\sqrt{\mu}\,
a^{\dagger}f(t)
-\sqrt{\mu}\,a\,\overline{f(t)}+\mu\,\mathrm{Re}^2\big(\langle a
\rangle_{t}\,\mathrm{e}^{-\mathrm{i}\phi(t)}\big)/2\right)
\widehat{\varphi}(t)\,\mathrm{d}t} \nonumber\\
&&+\mu a\,\mathrm{e}^{-\mathrm{i}\phi(t)}\mathrm{Re} \big(\langle
a\rangle_{t}\,\mathrm{e}^{-\mathrm{i}\phi(t)}\big)\,
\widehat{\varphi}(t)\,\mathrm{d}t+\left(\sqrt{\mu}\,a\,
\mathrm{e}^{-\mathrm{i}\phi(t)}- \sqrt{\mu}\,\mathrm{Re}
\big(\langle a\rangle_{t}\,\mathrm{e}^{-\mathrm{i}\phi(t)}
\big)\right)\, \widehat{\varphi}(t)\nonumber\\&&
\times\left(\mathrm{d}W(t)-2\sqrt{\mu}\,\mathrm{Re}
\big(\langle a\rangle_{t}\mathrm{e}^{-\mathrm{i}\phi(t)}
\mathrm{d}t)\right)\,,\;
\end{eqnarray}
%-------------------------------------------------------------------------
where $\langle a\rangle_{t}=\langle\widehat{\varphi}(t)|a\,
\widehat{\varphi}(t)\rangle$. And if the initial state of $\mathcal{S}$ is a
mixed one, $\widehat{\varrho}(0)=\varrho$, then the posterior
normalized density matrix $\widehat{\rho}(t)$ satisfies the
nonlinear filtering equation of the form
%-------------------------------------------------------------------------
\begin{eqnarray}\label{diff9}
\mathrm{d}\widehat{\rho}(t)&=& \left(-\frac{\mathrm{i}}
{\hbar}[H,\widehat{\rho}(t)]- \frac{\mu}{2}\{a^{\dagger}a,\widehat{\rho}
(t)\} +[\sqrt{\mu}a\overline{f(t)}- \sqrt{\mu}\,a^{\dagger}f(t),
\widehat{\rho}(t)]+\mu
a\widehat{\rho}(t)a^{\dagger}\right)\mathrm{d}t
\nonumber\\
&&
+\left(\sqrt{\mu}\,a\,\mathrm{e}^{-\mathrm{i}\phi(t)}\,\widehat{\rho}(t)
+\widehat{\rho}(t)  \sqrt{\mu}\,a^{\dagger}\,
\mathrm{e}^{\mathrm{i}\phi(t)}-2\,\sqrt{\mu}\,\mathrm{Re}
\big(\langle a\rangle_{t}\, \mathrm{e}^{-\mathrm{i}\phi(t)}
\big)\right)\nonumber\\
&&\times\left(\mathrm{d}W(t)-2\sqrt{\mu}\,\mathrm{Re}
\big(\langle a\rangle_{t}\mathrm{e}^{-\mathrm{i}\phi(t)}
\mathrm{d}t)\right)\,,
\end{eqnarray}
%-------------------------------------------------------------------------
where $\langle a\rangle_{t}=\mathrm{Tr}[\widehat{\rho}(t)a]$.
Eq. (\ref{diff9}) is consistent with the result in \cite{DSOSAB2011}.

\section{Posterior evolution of a squeezed coherent state}

We shall show that the coherent state survives the reduction of the
state following the registered trajectory: the solution to Eq.
(\ref{diff7}) for the initial coherent state $|\alpha_{0} \rangle$,
$\alpha_{0} \in \mathbb{C}$, can be written in the form
$|\widehat\psi (t) \rangle = l(t) |\alpha(t) \rangle$. Inserting
the predicted solution into Eq. (\ref{diff7}) and making use of the
property
%-------------------------------------------------------------------------------
\begin{equation}
a^{\dagger}|\alpha\rangle\;=\;\frac{\partial|\alpha\rangle}{\partial
\alpha}+\frac{1}{2}\overline{\alpha}|\alpha\rangle\,,
\end{equation}
%-------------------------------------------------------------------------------
on can write both sides of the equation in terms of linearly
independent vectors $|\alpha\rangle$ and $\frac{\partial
|\alpha\rangle}{\partial \alpha}$. Comparing the coefficients of
the corresponding vectors one gets the consistent system of
differential equations
%------------------------------------------------------------------------------
\begin{equation}
\mathrm{d}\alpha (t )\;=\; -\bigg(\mathrm{i}\omega +{\mu \over 2}
\bigg)\alpha (t)\mathrm{d}t-\sqrt{\mu}f(t)\mathrm{d}t\,,
\end{equation}
\begin{eqnarray}
\frac{\mathrm{d}l(t)}{l(t)}\;=\;-{\mathrm{i}\omega \over 2}
\mathrm{d}t-{\mu\over 2}|\alpha(t)|^{2}\mathrm{d}t-
\frac{3\sqrt{\mu}}{2}\alpha(t)\overline{f(t)}
\mathrm{d}t\nonumber\\
-\frac{\sqrt{\mu}}{2}\,\overline{\alpha(t)}f(t)\mathrm{d}t+ \sqrt
{\mu }\,\alpha (t)\mathrm{e}^{-\mathrm{i}\phi(t)} \,
\mathrm{d}W(t),
\end{eqnarray}
%-------------------------------------------------------------------------------
for the functions $\alpha(t)$ and $l(t)$ with the initial condition
$\alpha(0) = \alpha_{0}$, $l(0) = 1$. To solve the stochastic
equation for $l(t)$ one has to use the Ito rules and the formula
%-------------------------------------------------------------------------
\begin{equation}
\mathrm{d}\ln l(t)=\frac{1}{l(t)}\mathrm{d}l-
\frac{1}{2l^{2}(t)} \left(\mathrm{d}l(t)\right)^2\,.
\end{equation}
%-------------------------------------------------------------------------
Taking into account that $(\mathrm{d}l(t))^2=\mu\,\alpha^2(t)
\mathrm{e}^{-2\mathrm{i}\phi(t)} \, \mathrm{d}t$, we obtain the
analytical solution of Eq. (\ref{diff7}) for the initial coherent
state in the form
%------------------------------------------------------------------------------
\begin{eqnarray}\label{sol1}
|\widehat \psi_{\alpha_{0}} (t) \rangle \;=\;
\exp \left[-{\mathrm{i}\omega t\over 2}
-\sqrt{\mu}\int_{0}^{t}\left({\sqrt{\mu}\over 2}
|\alpha(t^{\prime})|^{2}+
\alpha(t^{\prime})\overline{f(t^{\prime})}\right)
\mathrm{d}t^{\prime}\right.\nonumber\\
-\sqrt{\mu}\int_{0}^{t}\left(\mathrm{Re}\,
(\alpha(t^{\prime})\overline{f(t^{\prime})})
+\frac{\sqrt{\mu}}{2}\alpha^2(t^{\prime})
\mathrm{e}^{-2\mathrm{i}\phi(t^{\prime})}
\right)\mathrm{d}t^{\prime}\nonumber\\
\left.+\sqrt {\mu }\int_{0}^{t}\alpha (t^{\prime})\,
\mathrm{e}^{-\mathrm{i}\phi(t^{\prime})} \, \mathrm{d}
W(t^{\prime})\right]|\alpha(t)\rangle\,,
\end{eqnarray}
%--------------------------------------------------------------------------
where the independent of the noise amplitude $\alpha(t)$ reads
%--------------------------------------------------------------------------
\begin{equation}\label{solvalpha1}
\alpha(t)\;=\;\alpha_{0}\,\mathrm{e}^{-\left(\mathrm{i}\omega +{\mu
\over 2} \right)t}-\sqrt{\mu} \int_{0}^{t}
\mathrm{e}^{-\left(\mathrm{i}\omega +{\mu \over 2}
\right)(t-t^{\prime})}f(t^{\prime})\mathrm{d}t^{\prime}\,.
\end{equation}
%--------------------------------------------------------------------------
Since any initial state can be represented in the basis of the coherent
states
%-------------------------------------------------------------------------
\begin{equation}\label{ini}
|\psi(0)\rangle\;=\;\frac{1}{\pi}\int \mathrm{d}^{2}\alpha_{0}\,
\langle\alpha_{0}| \psi(0) \rangle|\alpha_{0}\rangle\,,
\end{equation}
%----------------------------------------------------------------------
the linearity of Eq. (\ref{diff7}) allows us to
write the general solution as
%----------------------------------------------------------------------
\begin{equation}\label{gen}
|\widehat \psi(t)\rangle\;=\;\frac{1}{\pi}\int
\mathrm{d}^{2}\alpha_{0}\, \langle\alpha_{0}|\psi(0) \rangle
|\widehat \psi_{\alpha_{0}} (t) \rangle\,.
\end{equation}
%----------------------------------------------------------------------

The process of measurement changes the state of the system. When
the system interacts with the external Bose field but the signal is
not measured, the state in general becomes mixed. The nondemolition
observation of the Belavkin type gives the opportunity to get
information about the system and allowing to retain some properties
of the initial state, for example, its purity.

A coherent state is not the only one state invariant under the filtering
equation (\ref{diff7}). We shall prove that a squeezed coherent state
is preserved under the considered observation as well.
We will discuss the time development of the posterior wave function
of $\mathcal{S}$ for initial state of the form
%-------------------------------------------------------------------------
\begin{equation}\label{squ1}
|\widehat \psi (0) \rangle \;=\;S(\xi_{0} )D(\alpha_{0})|0
\rangle\;=\; S(\xi_{0} )|\alpha_{0} \rangle\;=\;|\xi_{0}
,\alpha_{0} \rangle\,,
\end{equation}
%-------------------------------------------------------------------------
where
%-------------------------------------------------------------------------
\begin{equation}
D(\alpha_{0})\;=\;\exp\big(\alpha_{0}
a^{\dagger}-\overline{\alpha}_{0} a\big)\,,\,\,\,\alpha_{0}\in\mathbb{C}\,,
\end{equation}
%-------------------------------------------------------------------------
and
%-------------------------------------------------------------------------
\begin{equation}
\!\,S(\xi_{0})= \exp \left( {1\over
2}\overline{\xi}_{0}a^{2}-{1\over 2} \xi_{0}
\big(a^{\dagger}\big)^{2}\right), \,\,\,
\xi_{0}=\mathrm{e}^{\mathrm{i}\theta_{0}} \varrho_{0} \in
\mathbb{C}.
\end{equation}
%------------------------------------------------------------------------
The state is generated by displacing the vacuum state and then by
squeezing. The amount of squeeze is described by
$\rho_{0}=|\xi_0|$ which is called the squeeze factor. Some details
of description and detection of squeezed states of light one can
find, for instance, in \cite{GK05}.

The method of computing a posterior dynamics for the initial
squeezed coherent state by making a simple ansatz like in the
previous case is cumbersome and laborious. Therefore, we shall use
a more efficient method which allows to avoid arduous computation
of a stochastic phase of posterior state.

We make use of the eigenvalue equation
%-------------------------------------------------------------------------
\begin{equation}\label{eigen2a}
S(\xi )a\,S^{\dagger}(\xi )|\xi ,\alpha \rangle \;=\;\alpha\, |\xi
, \alpha \rangle\,,
\end{equation}
%-------------------------------------------------------------------------
which can be readily fond from the definition (\ref{squ1}).
The operator expansion theorem allows one to check
that \cite{GK05}
%------------------------------------------------------------------------
\begin{equation}\label{eigen2b}
S(\xi )aS^{\dagger}(\xi )\;=\;a\,\Gamma _{1}+a^{\dagger}\,\Gamma
_{2}\,,
\end{equation}
%------------------------------------------------------------------------
where $\Gamma _{1}=\cosh \varrho$,
$\Gamma_{2}=\mathrm{e}^{\mathrm{i} \theta} \sinh \varrho$. Let us
notice that if the system $\mathcal{S}$ remains in the squeezed
coherent state at any time instant  $t\geq 0$, then
%------------------------------------------------------------------------
\begin{equation}\label{eigen3a}
S(\xi(t) )aS^{\dagger}(\xi(t) )\widehat{\psi}(t)\;=\;\left[a\,
\Gamma _{1} (t)+a^{\dagger}\,\Gamma
_{2}(t)\right]\widehat{\psi}(t)\,,
\end{equation}
%------------------------------------------------------------------------
and
%------------------------------------------------------------------------
\begin{eqnarray}\label{eigen3b}
\lefteqn{S(\xi(t+\mathrm{d}t) )aS^{\dagger}(\xi(t+\mathrm{d}t)
)\widehat{\psi} (t+\mathrm{d}t)} \nonumber \\
&& = \left[a\, \Gamma _{1}(t+\mathrm{d}t)+a^{\dagger}\, \Gamma
_{2}(t+\mathrm{d}t)\right]\widehat{\psi}(t+\mathrm{d}t)\,,
\end{eqnarray}
%------------------------------------------------------------------------
where $|\widehat\psi(t)\rangle=l(t)S(\xi(t))|\alpha (t) \rangle$,
have to be fulfilled. Eqs. (\ref{eigen3a}) and (\ref{eigen3b}) can
be reduced to a single condition of the form
%------------------------------------------------------------------------
\begin{eqnarray}\label{eigen3c}
&&\!\!\!\!\!\!\!\!\left[a \left(\Gamma
_{1}(t)\!+\!\mathrm{d}\Gamma _{1}(t) \right)\!+\! a^{\dagger}\,
\left( \Gamma _{2}(t)\!+\!\mathrm{d}\Gamma _{2}(t)\right)
\!-\!\alpha (t)\!-\!\mathrm{d}\alpha(t)\right] \mathrm{d}\widehat
\psi
(t)\nonumber\\
&&\!\!\!\!+\left(a\,\mathrm{d}\Gamma
_{1}(t)+a^{\dagger}\,\mathrm{d}\Gamma _{2}(t)-\mathrm{d}\alpha
(t)\right)\widehat \psi (t)\;=\;0\,.
\end{eqnarray}
%------------------------------------------------------------------------
Finally inserting of the increment $\mathrm{d}\widehat{\psi}(t)$
given by Eq.~(\ref{diff7}) into Eq.~(\ref{eigen3c}) we obtain the
set of the two differential equations:
%------------------------------------------------------------------------
\begin{eqnarray}\label{diff3a}
&&\!\!\!\!\!\!\!\!\!\Gamma_{2}(t)\left[ \!-\mathrm{\Gamma}_{1}(t)
\left(\mathrm{i}\omega+ \frac{\mu}{2}\right)\mathrm{d}t +\! \mu\,
\mathrm{e}^{-2\mathrm{i}\phi(t)}\Gamma_{2}(t)\mathrm{d}t+
\!\mathrm{d}\mathrm{\Gamma}_{1}(t)\right] \;
\nonumber\\
&&\!\!\!\!\!\!-\Gamma_{1}(t)\left[\mathrm{\Gamma_{2}(t)}\left(\mathrm{i}
\omega + \frac{\mu}{2}\right)\mathrm{d}t+
\mathrm{d}\mathrm{\Gamma}_{2}(t) \right]\;=\;0\,,
\end{eqnarray}
%------------------------------------------------------------------------
%------------------------------------------------------------------------
\begin{eqnarray}\label{diff3b}
&& \!\!\!\!\!\!\!\!\!\!\!\!
\alpha(t)\Gamma_{1}(t)\!\left[-\Gamma_{1}(t)
\left(\mathrm{i}\omega \!+\!
\frac{\mu}{2}\right)\mathrm{d}t\!+\!\! \mu\,
\mathrm{e}^{-2\mathrm{i}\phi(t)}\Gamma_{2}(t)\mathrm{d}t\!+
\!\mathrm{d} \Gamma_{1}(t)\right]
\nonumber\\
&&\!\!\!\!\!\!\!\!\!
\!-\!\alpha(t)\,\overline{\Gamma_{2}(t)}\left[\Gamma_{2}(t) \left(
\mathrm{i}\omega\!+\!\frac{\mu}{2}\right)\mathrm{d}t\!+\!
\mathrm{d}\Gamma_{2}(t)\right]\!-\! \sqrt{\mu}\,\Gamma_{1}(t)
f(t)\mathrm{d}t\\
&&\!\!\!\!\!\!\!\!\! -\sqrt{\mu}\,\Gamma_{2}(t)\overline{f(t)}\mathrm{d}t
-\sqrt{\mu}\,\Gamma_{2}(t)\mathrm{e}^{-\mathrm{i} \phi(t)}
\mathrm{d}W(t)-\mathrm{d}\alpha(t)\,=\,0\,\nonumber
\end{eqnarray}
%------------------------------------------------------------------------
with the initial condition: $\Gamma_{1}(0)=\cosh{\varrho_{0}}$,
$\Gamma_{2}(0)=\mathrm{e}^{\mathrm{i}\theta_{0}}\sinh \varrho_{0}$,
$\alpha(0)=\alpha_{0}$. The last step requires left-multiplying the
Eq. (\ref{eigen3c}) by $S^{\dagger}(\xi(t))$ and use of the
transformation
%--------------------------------------------------------------------------
\begin{equation}
S^{\dagger}(\xi )aS(\xi )\;=\;a\,\Gamma _{1}-a^{\dagger}\,\Gamma
_{2}\,,
\end{equation}
%--------------------------------------------------------------------------
which one can easily get from (\ref{eigen2b}).
The Eqs.~(\ref{diff3a}), (\ref{diff3b}) form a consistent set of equations
and this is completes the proof.

Eq.~(\ref{diff3a}) can be rewritten in terms of the function
$\mathrm{\Gamma}(t)=\mathrm{\Gamma}_{2}(t)/
\mathrm{\Gamma}_{1}(t)$ as
%------------------------------------------------------------------------
\begin{equation}\label{Riccati}
\frac{\mathrm{d}}{\mathrm{d}t}\mathrm{\Gamma}(t)\;=\;
-2\left(\mathrm{i} \omega+\frac{\mu}{2}\right)\mathrm{\Gamma}(t)+
\mu\, \mathrm{e}^{-2\mathrm{i}\phi(t)}\,\mathrm{\Gamma}^{2}(t)\,.
\end{equation}
%------------------------------------------------------------------------
The general solution to the Riccati differential equation
(\ref{Riccati}) reads
%------------------------------------------------------------------------
\begin{equation}
\mathrm{\Gamma}(t)\;=\;
\frac{\mathrm{\Gamma}(0)\mathrm{e}^{-(2\mathrm{i} \omega+\mu)t}}{1
-\mu\, \Gamma(0)\int\limits_{0}^{t} \mathrm{e}^{-\big(2\mathrm{i}
\omega+\mu\big)t^{\prime}-2\mathrm{i}\phi(t^{\prime})}\mathrm{d}
t^{\prime}}\,.
\end{equation}
%------------------------------------------------------------------------
In particular, for the phase $\phi(t)=\pi/2-\omega_{0} t $, one
gets
%------------------------------------------------------------------------
\begin{equation}
\mathrm{\Gamma}(t)\;=\; \frac{\big(2\mathrm{i}\omega-2\mathrm{i}
\omega_{0}+ \mu\big)\mathrm{\Gamma}(0)}{
\mathrm{e}^{(2\mathrm{i}\omega+\mu)t}
\big[2\mathrm{i}\omega-2\mathrm{i}\omega_{0}+
\mu(1+\Gamma(0))\big]-\mu\,
\Gamma(0)\mathrm{e}^{2\mathrm{i}\omega_{0} t}}\,.
\end{equation}
%-------------------------------------------------------------------------
The integration of (\ref{diff3b}) yields the stochastic amplitude
%-------------------------------------------------------------------------
\begin{eqnarray}\label{solforalpha}
&&\!\!\!\!\!\!\!\!\!\!\!\! \alpha (t)= \frac{1}
{\sqrt{1-\left|\Gamma(t)\right|^2}}
\left[\alpha_{0} \sqrt{1-\left|\Gamma(0)\right|^2}
\right. \nonumber \\
&&\!\!\!\!\!\!\!\!\! \left. \times \exp \left(-\mathrm{i} \omega t
-{\mu \over 2}
t+\mu\int_{0}^{t}\mathrm{e}^{-2\mathrm{i}\phi(t^{\prime})}
\Gamma(t^{\prime})\mathrm{d}t^{\prime}\right)
\right.\nonumber\\
&& \!\!\!\!\!\!\!\!\! -\sqrt {\mu}\,\int_{0}^{t}
\exp\left(-\left(\mathrm{i} \omega+\frac{\mu}{2}\right)(t-s)+
\mu\int_{s}^{t} \mathrm{e}^{-2\mathrm{i}\phi(t^{\prime})}
\Gamma(t^{\prime})\mathrm{d}t^{\prime}\right)\\
&& \!\!\!\!\!\!\!\!\!\left.\times\left(f(s)\mathrm{d}s+\Gamma(s)
\overline{f(s)}\mathrm{d}s+
\mathrm{e}^{-\mathrm{i}\phi(s)}
\Gamma(s)\mathrm{d}W(s)\right)\right]\,.\nonumber
\end{eqnarray}
%-------------------------------------------------------------------------

Therefore the posterior mean values of the optical quadratures $X=(a+a^{\dagger})/2$
and $Y=(a-a^{\dagger})/2\mathrm{i}$ for the initial squeezed coherent state
given as
%-------------------------------------------------------------------------
\begin{equation}
\langle X\rangle_{t}=\frac{\mathrm{Re} \left(\alpha(t)-
\overline{\alpha(t)}\mathrm{\Gamma}(t)
\right)}{\sqrt{1-|\mathrm{\Gamma}(t)|^2}}\,,
\end{equation}
%-------------------------------------------------------------------------
%-------------------------------------------------------------------------
\begin{equation}
\langle Y\rangle_{t}=\frac{\mathrm{Im} \left(\alpha(t)-
\overline{\alpha(t)} \mathrm{\Gamma}(t)\right)}{
\sqrt{1-|\mathrm{\Gamma}(t)|^2}}\,
\end{equation}
%-------------------------------------------------------------------------
depend on the measurement noise, whereas the uncertainties of
quadratures
%-------------------------------------------------------------------------
\begin{equation}
\label{deltax2} \Delta X(t)\;=\; \left(4 \mathrm{Re}
\kappa(t)\right)^{-1/2}\,,
\end{equation}
%------------------------------------------------------------------------
%------------------------------------------------------------------------
\begin{equation}
\label{deltay2} \Delta Y(t)\;=\;|\kappa(t)|\left(4 \mathrm{Re}
\kappa(t)\right)^{-1/2}\,,
\end{equation}
%------------------------------------------------------------------------
where
%------------------------------------------------------------------------
\begin{equation}
\kappa(t)\;=\; \frac{1+\mathrm{\Gamma}(t)}{1-\mathrm{\Gamma}(t)}\,
\end{equation}
%------------------------------------------------------------------------
remain deterministic. Moreover, the expressions for the mean value
of optical quadratures include the parameter of the initial coherent
state of Bose field, whereas the formulae for the uncertainties
are exactly the same as for the case when the Bose field is initially
in the vacuum state \cite{DSPLA2011}.

The time dependence of the uncertainties
$\Delta X$ and $\Delta Y$ has been illustrated by the parametric
plots presented in Fig.~2. They show the dynamics of $\Delta X$
and $\Delta Y$ as functions of the dimensionless time $\tau =
\omega t$ ($0 \leq \tau \leq 100$), for $\omega_{0}=0.05$, $\Gamma(0)=0.8$
and three values of $\mu$. The uncertainties oscillate approaching
the asymptotic values ($\Delta X=\Delta Y=\frac{1}{2}$).
The lager parameter $\mu$ the asymptotic values are reached faster.

In order to study the probability density of the output process
(\ref{measure}) one has to find the norm $\|\widehat{\psi}(t)\|=|l(t)|$.
One can check that
%-------------------------------------------------------------------------
\begin{equation}
\frac{\mathrm{d}|l(t)|^2}{|l(t)|^2}\;=\;  \frac{2\sqrt{\mu}\,\mathrm{Re}\,\left[\left(\alpha(t) -
\overline{\alpha(t)}\Gamma(t)\right)\mathrm{e}^{-\mathrm{i}\phi(t)}
\right]
}{\sqrt{1-|\mathrm{\Gamma}(t)|^2}}\mathrm{d}W(t)\,.
\end{equation}
%-------------------------------------------------------------------------
Hence
%-------------------------------------------------------------------------
\begin{eqnarray}
|l(t)|^{2}\;=\;
\exp\left\{2\sqrt{\mu}\int_{0}^{t}\frac{\mathrm{Re}\,
\left[\left(\alpha(t^{\prime}) -\overline{\alpha(t^{\prime})}
\Gamma(t^{\prime})\right)\mathrm{e}^{-\mathrm{i}\phi(t^{\prime})}
\right]}{\sqrt{1-|\mathrm{\Gamma}(t^{\prime})|^2}}
\mathrm{d}W(t^{\prime})
\right.\\
\left.-2\mu \int_{0}^{t}
\frac{\mathrm{Re}^{2}\left[\left(\alpha(t^{\prime})
-\overline{\alpha(t^{\prime})}\Gamma(t^{\prime})\right)
\mathrm{e}^{-\mathrm{i}\phi(t^{\prime})}\right]}
{1-|\mathrm{\Gamma}(t^{\prime})|^{2}} \mathrm{d}t^{\prime}\right\}\,.
\nonumber
\end{eqnarray}
%-------------------------------------------------------------------------
Let us stress that the formula (\ref{measure}) gives  the
probability of observed result $q^{t}\in\Omega^{t}$.

We should remark that, due to $\lim_{t\rightarrow \infty}\mathrm{\Gamma}(t)=0$, the
system asymptotically approaches the coherent state with the
amplitude
%-------------------------------------------------------------------------
\begin{equation}\label{asympto}
\alpha(t)\;=\;-\sqrt{\mu}\int_{0}^{t}
\mathrm{e}^{-(\mathrm{i}\omega+ \frac{\mu}{2})(t-t^{\prime})}
f(t^{\prime})\mathrm{d}t^{\prime}\,.
\end{equation}
%-------------------------------------------------------------------------
Hence, any memory about the initial condition is lost after a
transitional period. By (\ref{gen}) the asymptotic posterior state
for any initial state is the coherent state with the amplitude
(\ref{asympto}).

\section{Final remarks}

We have presented the derivation of the Belavkin filtering equation
for a single-mode field in a cavity interacting with the Bose field
(measuring apparatus) initially prepared in a coherent state. The
considered balanced heterodyne observation is intense and is
therefore considered as a diffusion one. In contrast to the case of
the measuring apparatus prepared initially in the vacuum state,
when the initial squeezed coherent state is preserved
\cite{DSPLA2011} and driven to the vacuum asymptotic state, in the
case studied in this paper the initial squeezed coherent state is
asymptotically driven to the coherent one. This asymptotic state
does not depend on the initial parameters of the initial state of
the field in the cavity. Moreover, we have proved that any initial
state described by a square integrable  wave function relaxes to
the coherent state, with the amplitude dependent on the coupling
constant and the initial coherent state of the apparatus. The
driving force from the field can control the system and drive its
state to the coherent one with the given amplitude.

\newpage

\noindent {\bf Figure captions}

\begin{center}
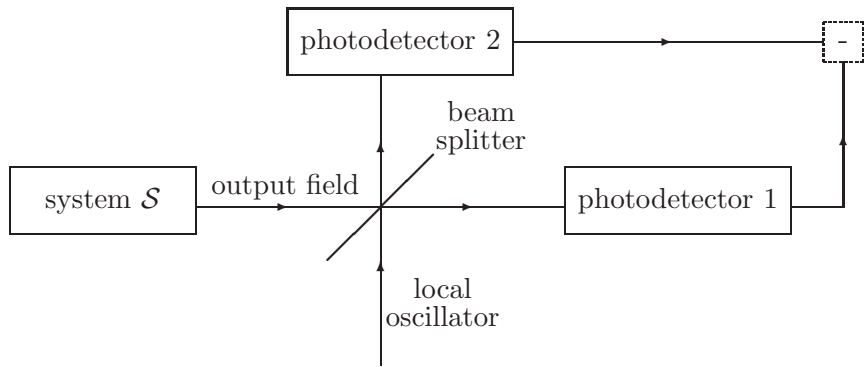
\begin{figure}[t]\label{Fig1}
\begin{picture}(200,170)(50,1)
\put(60,50){\framebox(70,25){\small{system $\mathcal{S}$}}}
\put(130,60){\vector(1,0){35}}
\put(165,60){\line(1,0){35}}
\put(136,65){\small{output field}}
{\thicklines
\put(180,40){\line(1,1){40}}}
\put(200,60){\vector(1,0){35}}
\put(235,60){\line(1,0){35}}
\put(270,50){\framebox(85,25){\small{photodetector 1}}}
\put(355,60){\line(1,0){20}}
\put(375,60){\vector(0,1){27.5}}
\put(375,87.5){\line(0,1){27.5}}
\put(200,60){\vector(0,1){25}}
\put(200,85){\line(0,1){25}}
\put(221,82){\shortstack{\small{beam}\\ \small{splitter}}}
\put(200,0){\vector(0,1){40}}
\put(200,40){\line(0,1){20}}
\put(165,110){\framebox(85,25){\small{photodetector 2}}}
\put(250,122.5){\vector(1,0){60.25}}
\put(301.25,122.5){\line(1,0){66.25}}
\put(367.5,115){\dashbox{1}(15,15){-}}
\put(202,15){\shortstack{\small{local}\\ \small{oscillator}}}
\end{picture}
\noindent\caption{\small Balanced heterodyne detection}
\end{figure}
\end{center}

\begin{figure}\label{Fig2}
\centering
\includegraphics[width=4.8cm,height=4.8cm, angle=270]{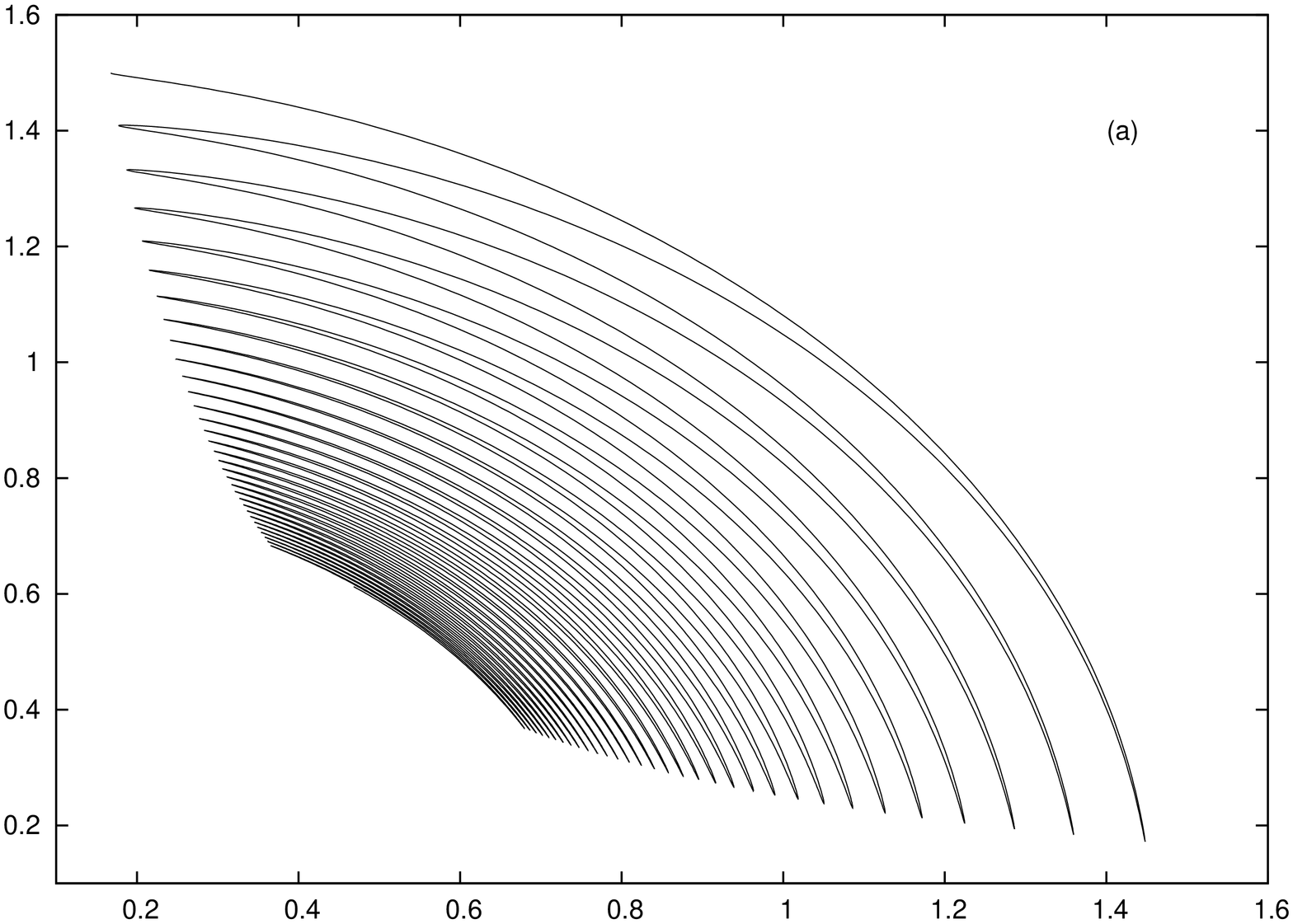}
\includegraphics[width=4.8cm,height=4.8cm,angle=270]{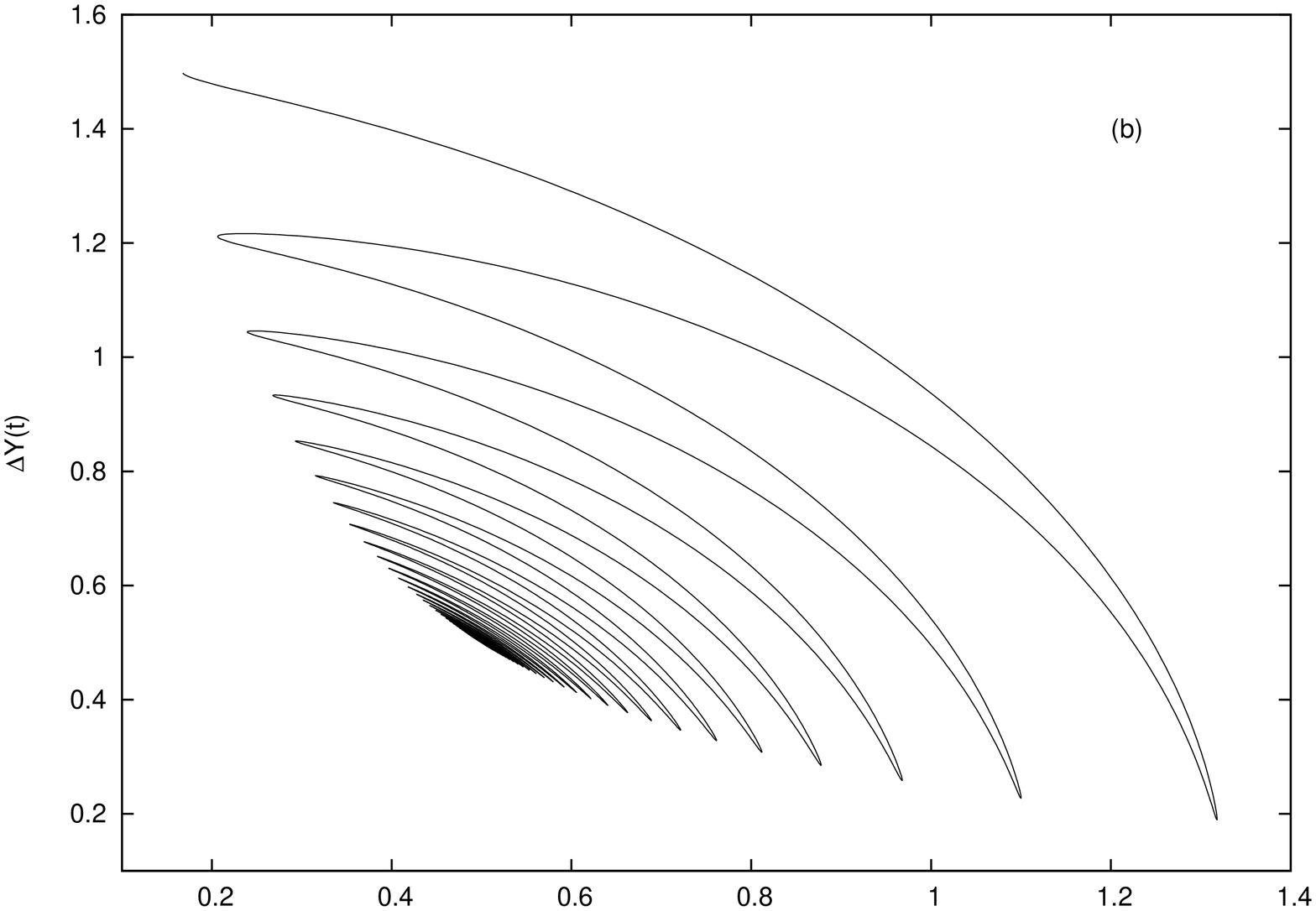}
\includegraphics[width=4.8cm,height=4.8cm,angle=270]{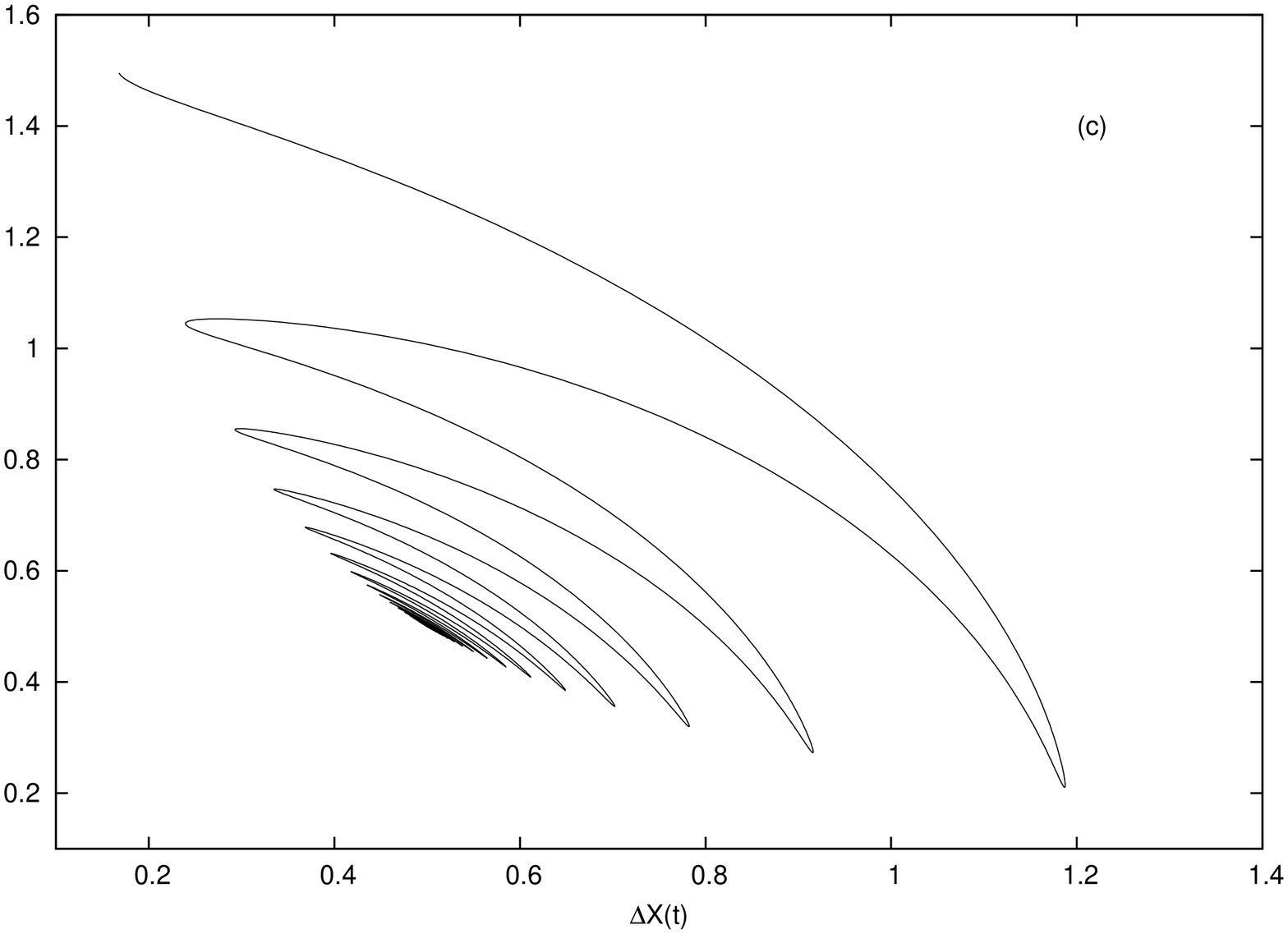}
\caption{The
dependence of $\Delta X$ and $\Delta Y$ on the dimensionless time
$\tau = \omega t$ is displayed for $\omega_{0}=0.05$, $\Gamma(0)=0.8$ and for three values of $\mu$: $0.01$ (a),
$0.04$ (b), and $0.08$ (c).}
\centering
\end{figure}
\end{document}